# Real-time measurement of laser induced photoconductivity in sol-gel derived Al doped ZnO thin films

F. Eskandari, M. Ranjbar*, P. Kameli and H. Salamati

Department of physics, Isfahan University of Technology, Isfahan, 84156-83111, Iran

**Abstract**

In this paper Al doped ZnO (AZO) thin films with 0, 3, 6 and 12 at. % Al concentration were prepared by sol-gel method on glass substrates. The deposited films were annealed at different temperatures of 300, 350, 400, 450 and 500 °C for 1 h in air. X-ray diffraction (XRD) showed wurtzite crystalline structure for the films annealed above 400 °C. The films were subsequently irradiated by beams of excimer (KrF, λ=248 nm) laser. The evolution of crystal structure, surface morphology and optical properties were studied using XRD, filed emission scanning electron microscope (FE-SEM) and UV-Vis spectrophotometer, respectively. Real-time measurement of electrical conductivity during laser irradiation showed a transient or persistent photoconductivity effect. The effect of laser energy on this photoconductivity was also investigated. Based on the observed photoluminescence (PL) and X-ray photoelectron spectroscopy (XPS), the observed photoconductivity effect was described.

## 1. Introduction

ZnO thin films has attractive characteristic features such as a direct band gap of 3.37 eV, large exciton binding energy of 60 meV and excellent chemical and thermal stability [1]. This features caused that ZnO thin films have a potential application in electronics, optoelectronics technology devices including displays, solar cells and sensors [2-4]. Several deposition techniques have been used to fabricate ZnO thin films, including sputtering [5], molecular beam epitaxy [6], pulsed laser deposition [7], spray pyrolysis [8] and the sol–gel deposition process [9]. Among them, sol–gel deposition process is not only simpler, but also offers the possibility of low cost preparing deposition of a large-area, at relatively low temperatures.

*Corresponding author. Tel: +98 31 33912375, fax: +98 31 33912376
Email Address: ranjbar@cc.iut.ac.ir (M. Ranjbar)

As-prepared ZnO films naturally shows *n*-type conductivity, due to the presence of zinc interstitials ($Zn_i$) and oxygen vacancies ($V_O$) [3, 10]. *N*-type conductivity of ZnO can also be achieved by doping with group III elements. These elements have one valence electron more than zinc, as a charge carrier. Because $Al^{+3}$ ions are substituted in $Zn^{+2}$ ions site, the extra free electrons increase the carrier concentration in the lattice. In addition, improvement of the crystalline structure and increasing the electron donors is considered to be pathways in order to reduce the electrical resistance of ZnO thin films [3]. Recently, laser radiation of thin films was used as an efficient method to improve the crystalline quality and increasing the electron donors in oxide semiconductor films [11-13]. Laser irradiation has several advantages over the other thermal annealing processes, including fast crystallization at room temperature, possibility of local crystallization, crystallization of thin films on low melting point substrates and increasing charge carriers through a photoconductivity effect. For example, Tsay et al. performed deposition ZnO thin films on glass at a temperature of 300 °C by sol–gel spin coating and used KrF excimer laser for annealing [14]. They showed the crystallinity levels and average crystallite size of excimer laser irradiated thin films were greater than those of thermally annealed thin films. Zhao et al. investigated the effect of KrF pulsed excimer laser irradiation on the structural, photoluminescence, and electrical properties and on the surface morphology of ZnO thin films [11]. They showed, the laser irradiated ZnO thin films exhibit a series of desirable properties compared to the as grown sample: UV emission is higher, resistivity is decreased by three orders of magnitude and the surface is flat and smooth. At the same time, it maintains a good epitaxial orientation and a wurtzite crystal structure. Tsang et al. fabricated transparent conducting AZO thin films using KrF excimer laser irradiation of sol–gel spin coated films [15]. They showed that reduction of electrical resistance is due to the improvements of in the crystalline quality and removing impurities. They also reported that the electrical and optical characteristics of AZO thin films were significantly improved by excimer laser irradiation. Most of reports have been done on the amorphous or weak crystalline films that annealed by excimer laser [16, 17]. So it seems interesting to study the effect of laser irradiation on other aspect of AZO films. In this study, we investigated the effects of excimer laser irradiation on amorphous and crystalline AZO thin films with different Al level, annealed at different temperatures. Additionally, the transient and persistent conductivity were investigated by real-time monitoring of conductivity during laser irradiation.

## 2. Experimental

The primary sol was prepared using zinc acetate dehydrate (ZnAc) ($Zn(C_2H_3O_2)_2.2H_2O$, Merck_99.99%), diethanolamine (DEA, Merck_99.99%), and isopropanol (iPrOH, Merck_99.5%) as precursor, stabilizing and the solvent agent, respectively. DEA was first dissolved in isopropanol. Then zinc acetate dihydrate was added under stirring, and heated for 1 h at 70 °C. Doping was done by adding a 0.2 M solution of aluminum nitrate in ethanol. The molar ratios of dopant in the solution, Al:Zn, were 0, 3, 6 and 12 at.%. Isopropanol was added to adjust the solution concentration to 0.5 mol/l of ZnAc. The molar ratio of DEA to ZnAc was maintained at 1.0. The solution was heated at 70 °C for 2 h to yield a clear and homogeneous solution, which used as the coating solution after cooling to room temperature. The coating was done 24 h after the solution was prepared. Spin coating technique was used to deposit the films on circular glass substrate at the speed of 3500 rpm for 30 s. The glass substrates were cleaned thoroughly by ethanol and deionized water in an ultrasonic bath and then dried. After each deposition stage the films were dried in air at 100 °C for 15 min. The films were then dried at 240 °C for 10 min in a furnace. The deposition process was repeated 8 times to get the desired thickness. As prepared films were then annealed in a furnace at 300–500 °C for 1 h in atmosphere, then the furnace turned off and the samples were kept in the furnace down to room temperature. The prepared films were irradiated by excimer laser (KrF, λ=248 nm). The crystalline structure of the films were characterized using X-ray diffractometer (Philips EXPERT MPD) with Cu-kα (λ = 0.154 nm) radiation. The morphology of the films was investigated using FE-SEM (Hitachi model S4460). Optical transmission spectra of the Zinc oxide films were investigated in the visible and near infrared (NIR) region, using a UV–vis spectrophotometer (model PerkinElmer lambda 25). The growth and decay of the photocurrents were measured between the two silver pads at a bias voltage of 10 V serried with a reference resistance of 11.5 kΩ. Room temperature photoluminescence (PL) spectra were recorded on a spectrometer (FP-750 model) at excitation wavelength of 320 nm of Xe lamp. Surface analysis of the samples was done by X-ray photoelectron spectroscopy (XPS)in an ESCA/AES system. The system is equipped with a concentric hemispherical analyzer (CHA, Specs model EA10 plus). For exciting the X-ray photoelectrons, an Al Kα line at 1486.6 eV was used. The energy scale was calibrated against the carbon binding energy (284.8 eV).

3. Results and discussion

*3.1.Crystalline structure*

Fig.1 shows XRD patterns of ZnO thin films on glass substrates annealed for 1 h in ambient air in the temperatures range of 300-500 °C. Diffraction peaks in the patterns of films annealed at 300 and 350 °C are weak or not observable. This indicates that the crystal structure of the sample annealed below 350 °C was not formed. As the temperature increases up to 500 °C, main diffraction peaks including (100), (002) and (101) gradually appears and their intensity increases with increasing the annealing temperature. The corresponding crystal structures of films annealed above 350 °C are hexagonal wurtzite (JCPDS, card no.01-075-0576). The film annealed at 500 °C has additional peaks, which may correspond to the granular structure of the film. The grain size of the ZnO films was calculated using the average grain size of three main (100), (002) and (101) peaks from the Scherrer's equation [18]. The calculated grain sizes were 22.8, 63.6, 75.1 nm for the sample annealed at the temperatures of 400, 450 and 500 °C, respectively. Obviously, the grain size has increased with annealing temperature. This effect had been also observed elsewhere [19, 20]. Since the crystalline ZnO films were obtained at 500 °C, in order to study the effect of Al doping on crystalline structure, AZO films were also annealed at 500 °C for 1 h. XRD patterns of AZO thin films with Al concentrations of 0, 3, 6 and 12 at. % are shown in Fig.2. As figure shows, the peaks intensities decrease by increasing doping content of aluminum, especially at sample with 12 at. % Al the peaks have entirely disappeared. This indicates, a large amount of Al dopants resulted in lattice disorder, which is associated with the stress generated by the difference in ion size between zinc and the aluminum [21]. This could occur because $Zn^{+2}$ cations are replaced with $Al^{+3}$ cations substitutionally in the lattice. At high Al content, the $Al^{+3}$ cations that are not able to substitute with secondary phase formation such as amorphous $Al_2O_3$ and their segregation in grain boundary also can be responsible for reducing crystallinity in AZO films [21-23]. According to Scherrer's formula the grains size of AZO thin film was 75, 35 and 25 nm for 0, 3 and 6 at. % Al, respectively. This shows grain size of AZO thin film was decreased with increasing Al concentration. It can be explained by the presence of the secondary phase such as $Al_2O_3$ inside ZnO grains which effectively pinned the movement of grain boundaries and prevents the grains growth and combination [21].

Fig.2 also compares XRD patterns of the ZnO annealed at 300 °C and AZO (0, 3, 6 and 12 at. % Al) films annealed at 500 °C with and without laser irradiation. The sample annealed at 300 °C includes an enhanced strong (002) peak around 2θ=34.5° after excimer laser irradiation, corresponding to preferential c-axis orientated wurtzite type crystalline structure. Winfield et al [17] and Kim et al [16] have been reported such phase formation for ZnO thin films annealed at 300 °C after laser irradiation. Therefore, laser irradiation causes the improving of crystalline properties of ZnO films. Beside a higher degree of crystallization could be observed at 500 ° (Fig.1 (e)), a phase deformation takes place for this case as the relative intensity of (002) peak increases under laser irradiation. Fig.2(b) shows the preferential re-crystallization along the c-axis orientation after laser irradiation. The average grain size decreases from 75.1 to 42.8 nm by laser irradiation of ZnO film annealed at 500 °C. The crystalline structures of AZO films undergo improving after laser irradiation as diffraction peaks slightly enhance for 3 and 6 at. % Al and very weak (101) peak appears at 36.4° for sample with 12 at. % Al. Indeed the organics residuals and acetate groups which remained from the sol-gel process exist in the as deposited films and are not completely desorbed form the composition even at 300 °C, so samples are amorphous. Pulses of high power excimer laser, which are effectively absorbed by the wide optical gap films, promote residual removal and partially crystallization. However, amorphous $Al_2O_3$ phase in AZO film with high concentration Al such as sample with 12 at. % Al are more stable and needs high enthalpy to remove.

*3.2. Morphology*

To understand the surface morphology evolution of ZnO and AZO films induced by excimer laser irradiation, FE-SEM was used (Fig.3). The FE-SEM micrographs of un-doped ZnO thin films annealed at 300 and 500 °C before laser irradiation show that the surface morphology and grain size varies with annealing temperature, i.e the grains develop by annealing at 500 °C. The average grain size was measured to be ~20 nm for sample annealed at 300 °C which was increased to ~70 nm for annealing temperature of 500 °C, in agreement with the XRD results. Adding Al has a significant impact on the morphology; even at an annealing temperature of 500 °C a large reduction in grain size is observable. Moreover, the grain size becomes smaller with increasing the doping level. The average grains size has decreased from ~70 (0 at. % Al) to ~15 nm (12 at. % Al). Regarding to XRD and FE-SEM results it seems that the doping is more difficult for the nanometer crystallites because more amorphous phase is provided to the grains boundary by increasing of the Al contents level but on the contrary, formation of more amorphous $Al_2O_3$ deteriorates it faster.

As can be seen, as a result of interaction of laser light with pure ZnO and AZO thin films surface, in all cases, the surface uniformity is lost and several cracks and voids are created. The voids are significant for pure ZnO film annealed at 300 °C, because the acetate groups compound desorption by laser irradiation results in films shrinkage and voids formation. Fig.3 shows that except the ZnO film annealed at 500 °C, the grain size slightly increases by laser irradiation. Also, the inter-granular micro-cracks maybe formed due to local non-uniform tensile stress induced by the rapid cooling rate after irradiation with excimer laser. Consequently, the grains with smaller size become larger and the growth orientation becomes strong after laser irradiation, thus leading to an improvement in the crystallinity of AZO. However, the grain size of pure ZnO thin film annealed at 500 °C become smaller after laser irradiation. This behavior can be explained by the fact that in pure ZnO annealed at 500 °C, surface to volume ratio is low, so with absorption of laser energy particles do not interact effectively, and instead the sticks are fragmented.

*3.3. Optical properties*

The effect of laser irradiation on optical properties of different type of sample was investigated for the UV-vis range. The results are presented in Fig.4(a-c). Typical optical transmission spectra of ZnO films at different annealing temperatures are shown in Fig.4(a). The transmission spectra exhibit sharp absorption edges around 380 nm indicating the general absorption behavior of thin zinc oxide films near the UV region. This behavior was observed for all type of AZO samples too. The optical band gaps, $E_g$, were determined by using the Tauc method [24]:

$$(\alpha h\nu) = A(h\nu - E_g)^{1/2} \qquad (1)$$

where α is the absorption coefficient, hν is the photon energy, A is a constant and $E_g$ is the optical band gap. The optical band gaps of sample were determined by extrapolation of the straight section to the energy axis of the plot of $(\alpha h\nu)^2$ versus photon energy. Inset of Fig.4(a) show typical extrapolation for ZnO films. The effect of laser irradiation on the optical band gap and average optical transmission of the ZnO thin films as a function of annealing temperature are shown in Fig.4(b). From the figure, when the annealing temperature increases from 300 up to 500 °C, the average optical transmission of the ZnO thin films increases slightly and laser irradiation has no impact on it. The optical band gap of non-irradiated films decreases from 3.38 to 3.18 eV by increasing annealing temperature. After laser the optical band gap decreases with annealing temperature while with a shift to higher energies. This may be due to the decreased optical scattering caused by the decrease in surface/volume ratio of grains owing to the increasing of size, improving crystallinity and decreasing of defects at elevated annealing temperature [25]. The optical band gap of crystalline ZnO thin films was reported to be lower than that of its amorphous phase[26].

The effect of Al doping level, before and after laser irradiation, on band gap and average optical transmission are shown in Fig.4(c). As can be seen with and without laser treatment, the optical transmission slightly reduces by increasing Al concentration. It has been attributed to the occupation of interstitial sites by Al atoms and formation of secondary phases such as $Al_2O_3$ in AZO thin films, which cause scattering the incident light [27]. Overall, it can be stated that there is no remarkable change in optical transmission with Al concentration and all samples have transparency above 80%. However, optical band gap increases by increasing Al concentration up to 6%. For non-irradiated sample the band gap reduces at 12% Al while for irradiated one it remains nearly unchanged. As can be seen, optical band gap increases by laser irradiation. These observation might be due to the Burstein-Moss effect which described the blue-shifting of the absorption edge of a semiconductor with an increasing carrier concentration [28].

*3.4. Electrical properties*

Fig.5 shows real-time of voltage monitoring during excimer laser irradiation of constant resistance $R_c$, connected in series with AZO samples ($R_s$) involving 10 V circuit bias voltage (inset shows the circuit). So, in the voltage curves, increasing the $V_{Rc}$ means the decreasing of the $R_s$ and vice versa. Each voltage curve involves five different raising-decaying cycles. At first, 50 laser pulses with a fluence of 190 mJ/cm² and repetition rate of 1 Hz were delivered to the AZO films surface. In this case, we choose a low repetition rate to compare the transient conductivity of samples against laser irradiation. This creates also an electrical pre-activation in the films. As can be seen in Fig.5, as the laser irradiation begins, an increase in the conductivity is observed for all types of the films. The next $2^{th}$-$5^{th}$ cycles involve 3000 pulses per cycle, repetition rate of 20 Hz and different fluences of 190, 130, 60 and 190 mJ/cm² for $2^{th}$, $3^{th}$, $4^{th}$ and $5^{th}$ one, respectively. The laser fluences at $2^{th}$ and $5^{th}$ cycles are the same to allow examination the stability of persistent photoconductivity over the laser irradiation time. At laser irradiation onset a rising conductivity begins to develop and reaches a saturation point. Then, after turning the laser off, conductivity decays exponentially to zero or nonzero values. The nonzero values clearly point out a persistent photoconductivity effect could be occurred in ZnO films by laser irradiation. The saturated and persistent conductivity as well as recovery times differ for different samples. In the samples with more persistent conductivity, the recovery times are also longer. Besides, dependence of saturated photoconductivity to laser energy is different for different types of samples. In the following we try to discuss the behavior of each sample in detail.

For ZnO (0 at. % Al) thin films, by increasing annealing temperature the laser-induced photoconductivity in the $1^{th}$-$5^{th}$ cycles increases. In the annealing temperature below 450 °C, the conductivity is lower at lower laser fluences and in $5^{th}$ cycle it reaches to values lower than that of $2^{th}$ with the same laser energy. Thus it seems the photocondoctivity of films changes within irradiation time. Also, for annealing temperature below 450 °C the laser-induced conductivity decays rapidly and vanishes within a short time. While for sample annealed at 500 °C, it decays to nonzero saturated values at longer recovery times. In this case, the saturated conductivity is almost the same for different laser energies.

For AZO thin films with 3 at. % Al the rising of conductivity at the first step increases with annealing temperature. Moreover, the saturated conductivities at the $2^{th}$-$5^{th}$ steps present an enhancement at 450 and 500 °C. Below 450 °C the conductivity varies by laser energy while above 450 °C changing the laser energy doesn't affect the saturated conductivity and its values are almost the same at all steps. Besides, at all annealing temperatures, not only a persistent conductivity exists but also it increases with annealing temperature. Comparing to 0 at. % Al doped films, at same annealing temperatures, the saturated conductivity and the recovery times of 3 at. % Al sample are increased and the sensitivity to laser energy is decreased.

By increasing Al concentration up to 6 at.%, one can observe improvement of first steps photoconductivity at higher annealing temperatures, similar behavior of 0 and 3 at. % Al samples. The saturated conductivities are higher than those of 0 at. % Al and are comparable with 3 at. % Al doped films. Sensitivity to incident energy is also more pronounced for annealing temperatures below 350 °C while for 450 and 500 °C laser energy doesn't affect photoconductivity saturation level. There is an increase in persistent conductivity like to 3 at. % Al (at 500 °C) for both 450 and 500 °C.

For 12 at. % Al-doped films, there is a weak transient photoconductivity at 300°C at all irradiating steps with strong sensitivity against laser energy. By increasing the annealing temperature, the photoconductivity saturating level increases and the sensitivity to laser energy becomes weak. The photoconductivity also becomes remarkably stable with increasing temperature especially the persistent conductivity at 500 °C is comparable to 3 and 6 at. % Al-doped samples.

Comparing the real-time curves suggest that at 500 °C all types of samples and at 450 °C sample 6 at. % Al, the persistent photoconductivity effect is pronounces. Laser induced photoconductivity effect can be explained by improvement of crystallite properties of films, creating electron-holes and electron donors such as oxygen vacancies or zinc interstitials (according to the results of the PL and XPS in section of 3.5 and 3.6). It also has been shown elsewhere that the electron donors are increased after laser irradiation [11].

In the transient photoconductivity effect, it is presumed that this is due to recombination of the electron-holes and filling oxygen vacancies by oxygen molecules in the air. Additionally, the instability of saturated conductivity over the laser irradiation time for low annealing temperatures can be explained by improvement of crystalline structure as an irreversible variation, especially occurs for amorphous structures.

*3.5. Photoluminescence*

Fig.6 shows the typical room-temperature photoluminescence spectra of the ZnO and AZO (6 at. % Al) samples annealed at 500 °C before and after laser irradiation. The ZnO emission is generally classified into two categories. One is the ultraviolet emission of the near band edge (NBE) in the UV region related to the direct band gap transition and/or free-exciton recombination. The other region is the deep-level (DL) emission in the visible range originated from defects, such as oxygen vacancies or zinc interstitials [29-32]. From Fig.6, it can be seen that after laser irradiation the UV emission peak of ZnO sample annealed at 500 °C undergoes shifting from 387 (3.20) to 385 nm (3.22 eV), while its intensity remains almost constant. The intensity of UV emission peak in AZO sample rise after laser irradiation and shifted from 371(3.34 eV) to 380 nm (3.26 eV). The increasing in the intensity of UV emission peak is connected with the improving of the degree of c-axis orientation and crystalline quality of ZnO and AZO thin films [16, 33]. The shifts of the position of the UV emission after laser irradiation might be attributed to the change in the band gap (Fig.4). In the visible region of ZnO PL spectrum, two emission peaks appear around 447 (2.77 eV) and 497 nm (2.49 eV) after laser irradiation. Moreover, there is an emission peak at 487 nm (2.5 eV) for non-irradiated sample with 6 at. % Al. The laser irradiation of this sample causes appearing a new emission peak at 442 nm (2.8 eV) which is attributed to the zinc interstitial ($Zn_i$) [34, 35]. The origin of blue emission peak around 497 and 478 nm is oxygen vacancies ($V_O$) [36]. Generally, PL spectra indicate that for ZnO sample, $Zn_i$ and $V_O$ states increase after laser irradiation. However, for AZO film there is some $V_O$ before laser irradiation, and the amount of $Zn_i$ increases after laser irradiation.

*3.6. Surface chemical analysis by XPS*

In order to investigate the effect of laser irradiation on surface chemical composition, the surface chemical states of atomic oxygen and zinc in AZO films (6 at. % Al), was compared before and after laser irradiation by XPS. We expected that besides the improvement of crystallite properties, the variation of chemical bonding of ZnO thin films after laser irradiation influence the electrical properties, mainly due to an increase of oxygen vacancies or zinc interstitial that act as *n*-type dopants. Fig.7 shows the XPS survey scan spectra of AZO (6 at. % Al) thin film annealed at 500 °C before and after laser irradiation. The photoelectron peaks of the main elements, Zn, O and C are observable at around 1022, 532 and 285 eV, respectively. The presence of carbon can be explained by surface contamination and by the organic residues resulting from acetate group absorption during the annealing and laser irradiation processes. To study the bonding states of Zn and O in details, high resolution scan spectra of O1s and $Zn2p_{3/2}$ peaks were used (Fig.8).The asymmetric O1s peak on the surface can be consistently fitted by three nearly Gaussian binding energy components centered at 530.5, 531.8 and 532.8 eV which are denoted $O_1$, $O_2$ and $O_3$, respectively. The peak $O_1$ is attributed to $O^{-2}$ ions on the wurtzite structure of ZnO that surrounded by Zn atoms with their full complement of nearest neighbor $O^{-2}$ ions [37]. In other words, the intensity of this component is the measure of the amount of oxygen atoms in a fully oxidized stoichiometric surrounding. Therefore, the $O_1$ peak at the O1s spectrum can be attributed to the Zn–O bonds. The medium binding energy component ($O_2$) at 531.8 eV is usually attributed to presence of loosely bound $O^{-2}$ ions that are in oxygen deficient regions within the ZnO matrix [37]. As a result, changes in the intensity of this component may be in connection with the variations in the concentration of the oxygen vacancies ($V_O$). The high binding energy component located at 532.8 eV ($O_3$) is usually attributed to the presence of loosely bound oxygen on the surface of AZO film, belonging to a specific species, e.g., –CO , adsorbed $H_2O$ or adsorbed $O_2$. The spectra of symmetric $Zn2p_{3/2}$ peak is also shown in Fig.8. To calculate the relative elemental composition of Zn versus O, the peaks of Zn $2p_{3/2}$ and O1s XPS are integrated, and the areas are divided by their corresponding sensitivity factors (4.8 for Zn2p3/2 and 0.66 for O1s). The calculated Zn/O ratio is about 0.7. The ratio Zn/O is the same before and after laser irradiation. The low ratio of oxygen to zinc indicates oxygen vacancies are present on the surface before and after laser irradiation in AZO (6 at. % Al) thin film. It was also predicted according to PL spectra. PL spectra show that a little oxygen vacancies have been produced after laser irradiation. While, interstitials Zn are the result of laser interactions with the surface of samples. This is in agreement with the XPS results that implies oxygen vacancies remains unchanged upon laser irradiation. Therefore we can conclude that Zn interstitials are a main reason for laser induced photoconductivity effect of ZnO and AZO films.

4. **Conclusions**

AZO/glass thin films with 0-12 at. % Al have been prepared by sol–gel processes. XRD spectra exhibited hexagonal wurtzite structure for ZnO thin film annealed at above 400 °C. Degree of crystallinity properties was observed to decrease with increasing Al concentration. We have demonstrated that KrF excimer laser irradiation can significantly improve the crystallinity level. Real-time measurement of electrical conductivity during laser irradiation showed a transient photocoductivity effect for 0 at. % Al and annealing temperature below 450 °C and persistent conductivity for 500 °C. Moreover, a persistent conductivity was observed for almost all the AZO films. For each sample types (0-12 at. % Al) the persistent photoconductivity was higher at 500 °C while the sample with 6% Al showed a high conductivity even at 450 °C. PL spectra showed that interstitial Zn ($Zn_i$) and a little amount of oxygen vacancies are produced within laser irradiation. This was in agreement with the XPS and results that imply oxygen vacancies remain unchanged upon laser irradiation. Generally, increasing in the conductivity of samples after laser irradiation was attributed to the production of electrons donor (oxygen vacancies and interstitials Zn) and improvement of crystallite structure of AZO thin films.

**Figure captions**

Fig. 1 XRD patterns of the pure ZnO thin film annealed at (a) 300, (b) 350, (c) 400, (d) 450, and (e) 500 °C, for 1h.

Fig. 2 XRD patterns of the pure ZnO thin films annealed at 300 and 500 °C and AZO thin films with different Al concentrations annealed at 500 °C before and after laser irradiation

Fig. 3 FE-SEM micrograph of pure ZnO thin films annealed at (a) 300, (b) 500 °C and AZO thin films with (c) 3, (d) 6 and (e) 12 at %.Al annealed at 500 °C, before (left) and after (right) laser irradiation.

Fig. 4 (a) Optical transmittance spectra of pure ZnO annealed at different temperatures, variation of optical band gap and average transmission as a function of (b) annealing temperature (c) Al concentration.

Fig. 5 Real-time curves of voltage of constant resistance, $R_c$, during excimer laser irradiation on AZO (0-12 at. % Al) thin films annealed at different temperatures.

Fig. 6 Photoluminescence emission spectra of pure ZnO annealed at 500 °C (a) before, (b) after laser irradiation and AZO (6 at.% Al) annealed at 500 °C (c) before and (d) after laser irradiation.

Fig. 7 XPS survey scan spectrum of AZO thin film with 6 at. % Al, annealed at 500 °C (a) before (b) after laser irradiation.

Fig. 8 High resolution XPS spectra of O1s and $Zn2p_{3/2}$ of AZO thin film with 6 at. % Al, annealed at 500 °C, before and after laser irradiation

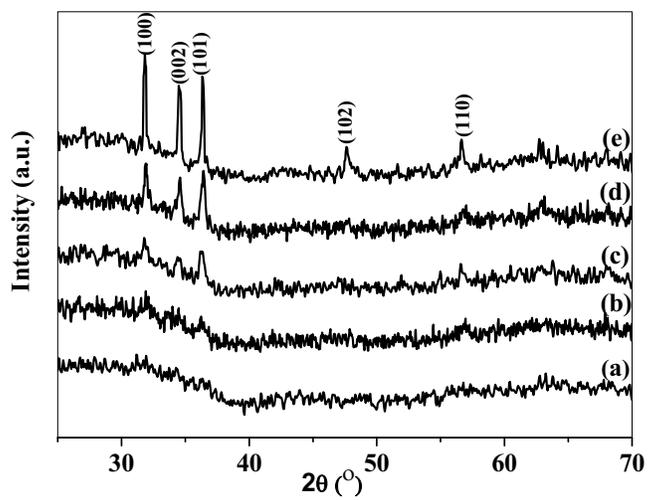

Fig. 1

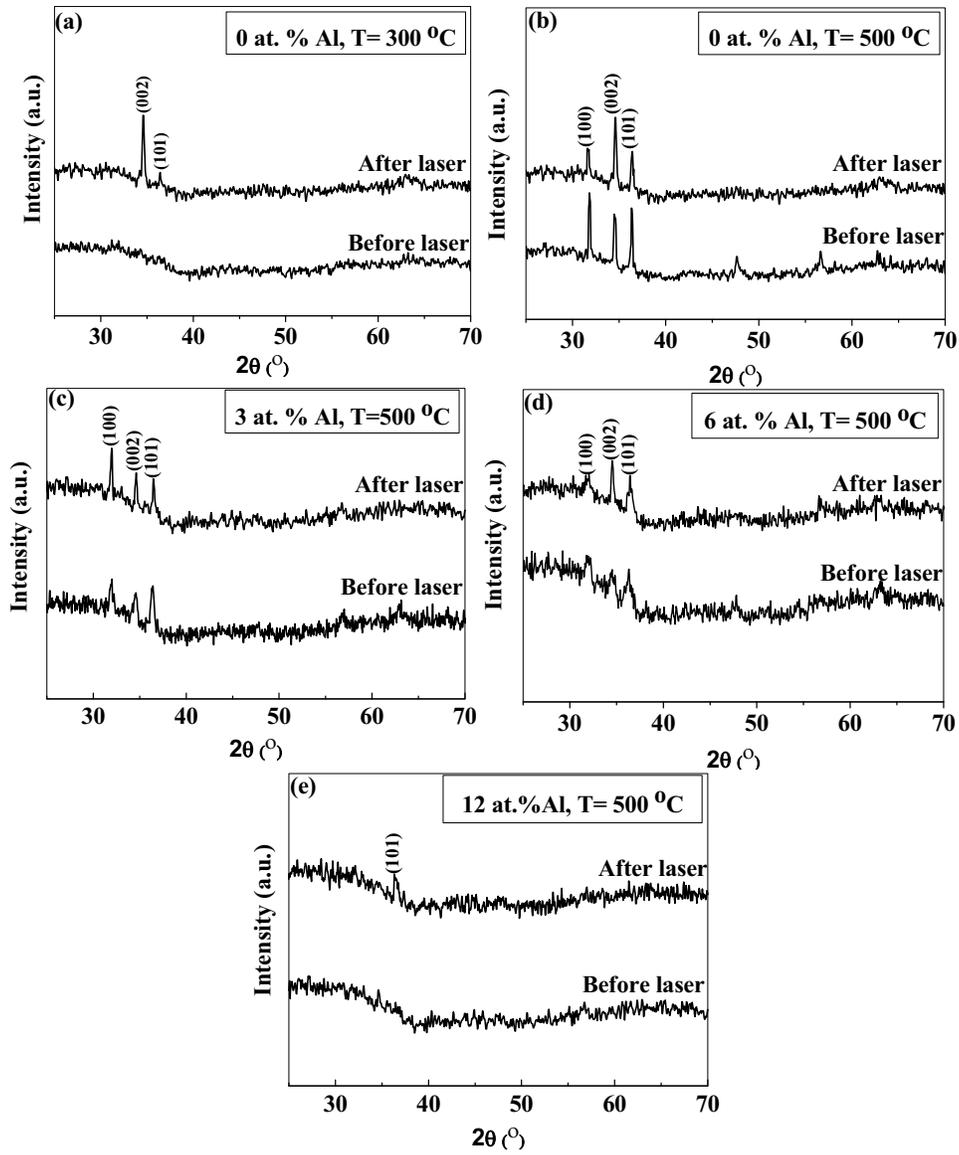

Fig. 2

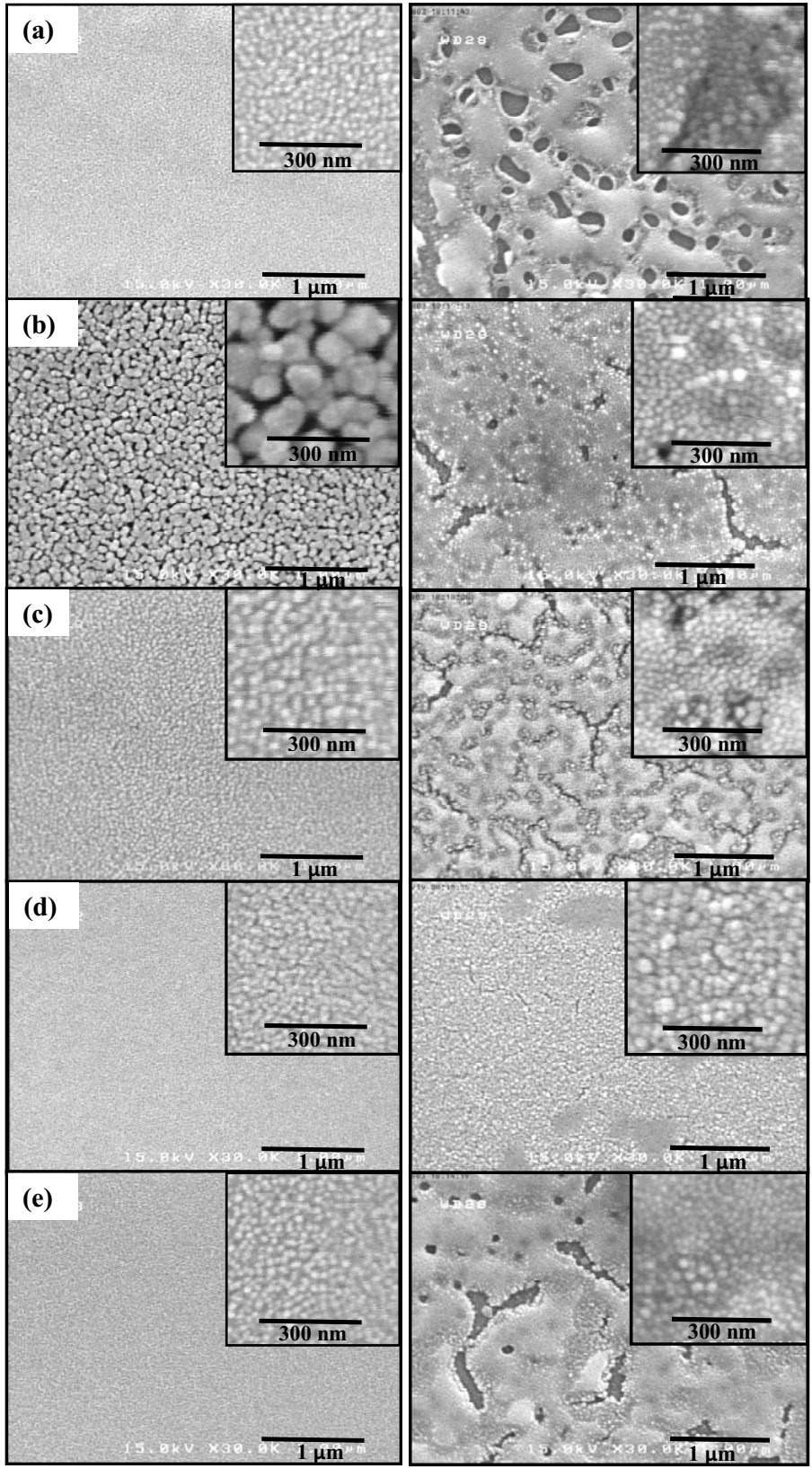

Fig. 3

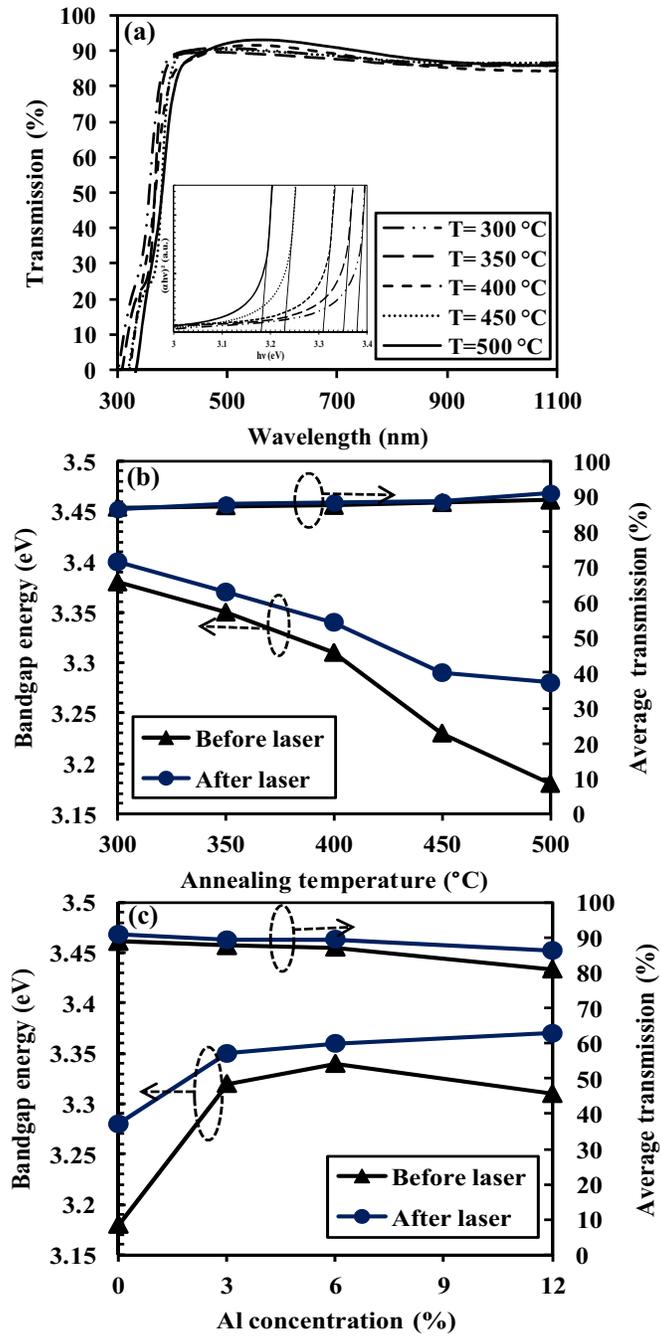

Fig. 4

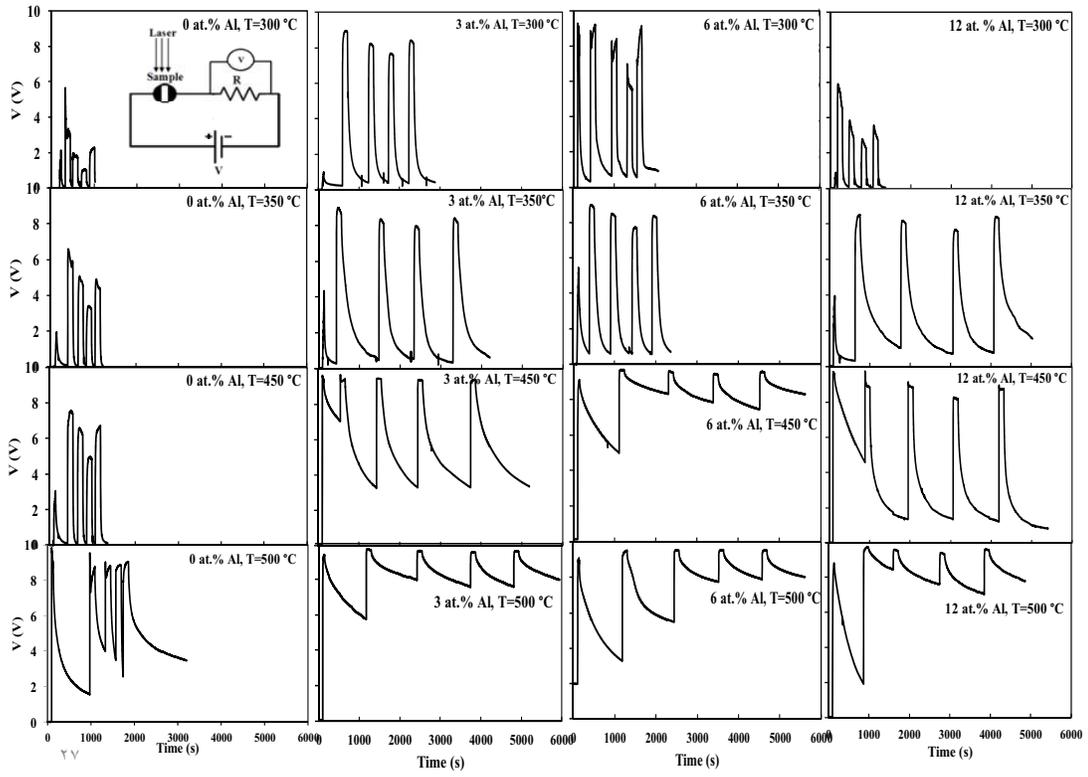

Fig. 5

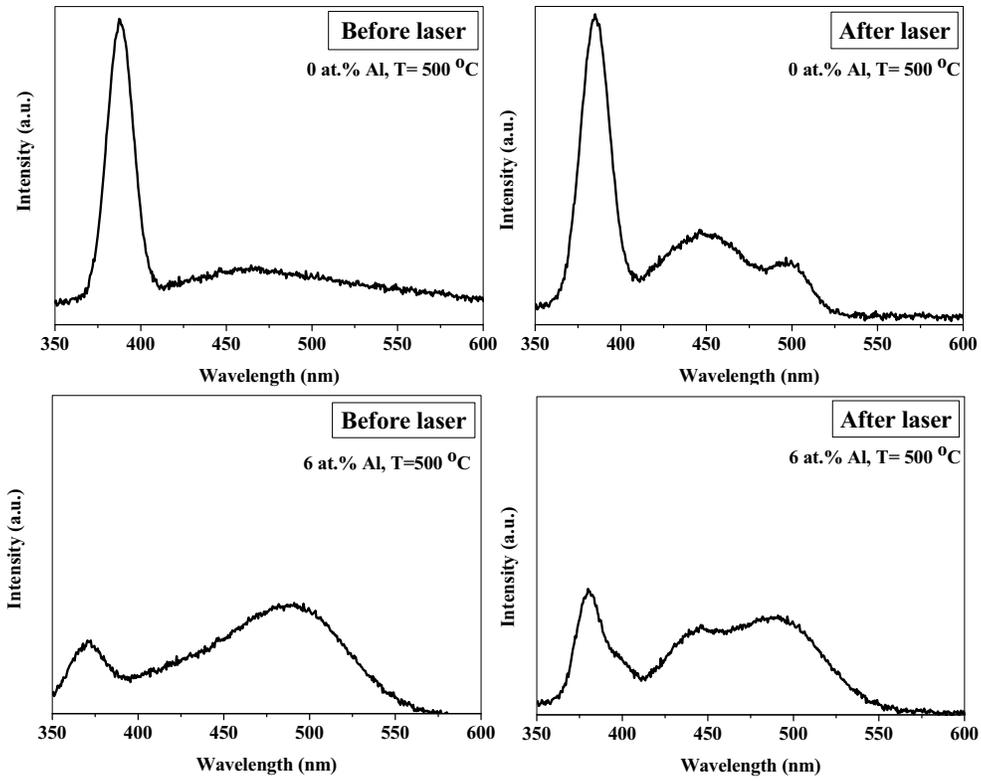

Fig. 6

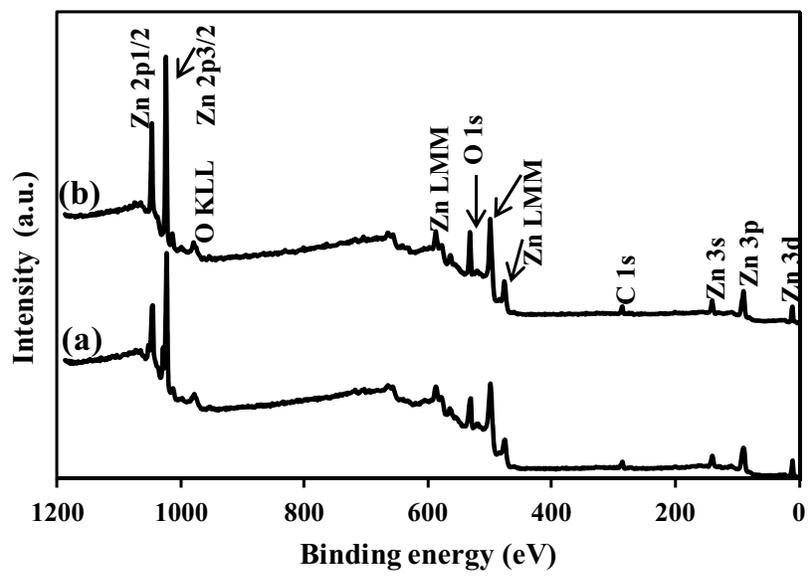

Fig. 7

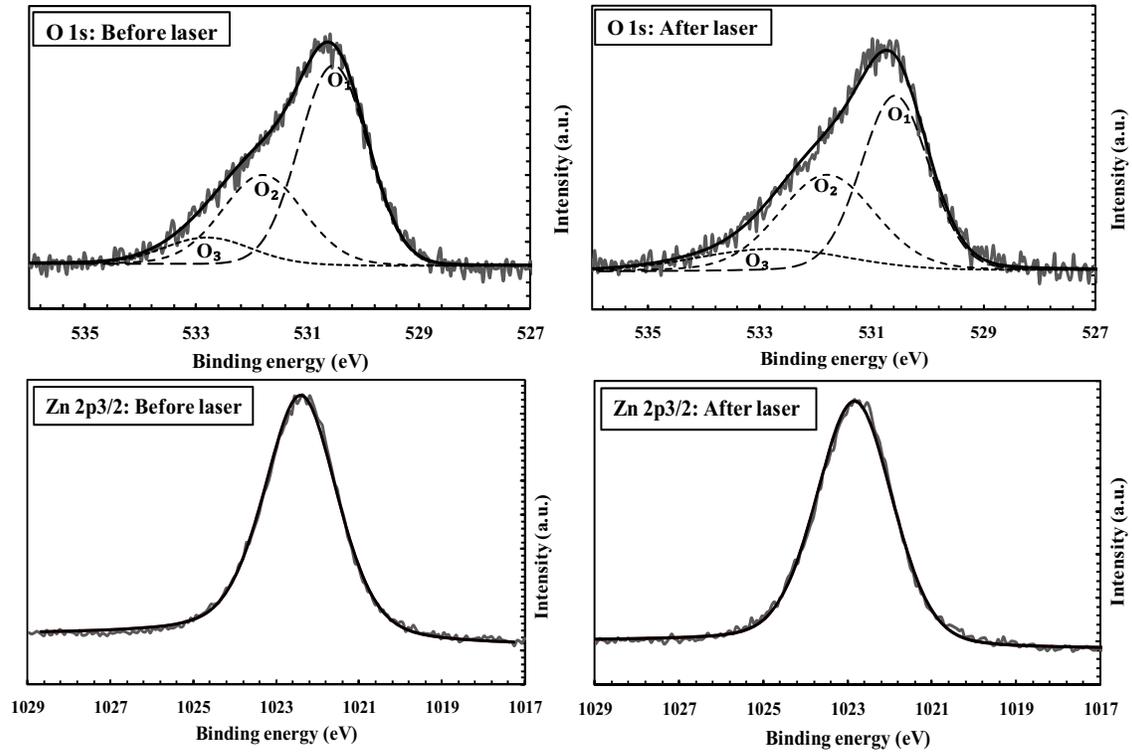

Fig. 8